\documentclass{ieeeaccess}

\usepackage{algorithmic}
\usepackage{graphicx}
\usepackage{amsmath,amssymb,amsfonts}
\usepackage{algorithmic}
\usepackage{graphicx}
\usepackage{textcomp}
\usepackage[sorting=none]{biblatex}  
\begin{document}

\history{Date of publication xxxx 00, 0000, date of current version xxxx 00, 0000.}
\doi{10.1109/ACCESS.2017.DOI}


\address[1]{Department of Computing Science of University of Alberta (e-mail ziyu18@ualberta.ca}
\address[2]{Department of Computing Science of University of Alberta (e-mail: fangyan1@ualberta.ca)}
\address[3]{Department of Computing Science of University of Alberta (e-mail:xguan3@ualberta.ca}

\tfootnote{This is a report for 3D Point Cloud Reconstruciton }

\markboth
{Author \headeretal: Preparation of Papers for IEEE TRANSACTIONS and JOURNALS}
{Author \headeretal: Preparation of Papers for IEEE TRANSACTIONS and JOURNALS}

\corresp{Corresponding author: Anup Basu. (e-mail: basu@ualberta.ca)}

\title{3D Point Cloud Reconstruction And SLAM As An Input }
\author{\uppercase{Ziyu Li},\uppercase{FangYang Ye}, \uppercase{Xinran Guan}
}
\address{Department of Computing Science, University of Alberta}


\begin{abstract}

To handle the different types of surface reconstruction tasks, we have replicated as well as modified a few of reconstruction methods and have made comparisons between the traditional method and data-driven method for reconstruction the surface of an object with dense point cloud as input. On top of that, we proposed a system using tightly-coupled SLAM as an input to generate deskewed point cloud and odometry and a Truncated Signed Distance Function based Surface Reconstruction Library. To get higher accuracy, IMU(Inertial Measurement Unit) pre-integration and pose graph optimization are conduct in the SLAM part. With the help of the Robot Operating System, we could build a system containing those two parts, which can conduct a real-time outdoor surface reconstruction.
\end{abstract}


\titlepgskip=-15pt
\maketitle

\section{Introduction}
\label{sec:introduction}

\PARstart{S}{urface} Reconstruction has been widely used in different fields. This Technology has been used to generate mesh of objects in a different scale, for example toys, the indoor scene, as well as outdoor environment. The reconstructed meshes can be utilized in many applications such as 3D printing, object volume estimation \cite{Buttress}, VR/AR scene modeling, indoor decoration, autonomous driving, remote sensing and mapping.

Different types of devices can be used to conduct the surface Reconstruction task and collect point cloud as an input, for example Monocular camera, stereo camera, RGB-D camera, and LiDAR. 


Compared with other devices, LiDAR can get directly get the point cloud of objects in a long range Therefore, what we mainly focus on is the surface reconstruction using point cloud collected by LiDAR in our project. After collecting point clouds from scenes and objects as inputs, we utilize different surface reconstruction algorithms to hanle different kinds of the input point cloud data. Since our final goal is to get the mesh of an object or a scene ,we need to define function to express the surface of the objects in implict or explicit way. Topological techniques play a crucial role while defining those functions. Moreover, researchers have been studying methodologies using discrete spatial data to construct the surface of objects or to build outdoor and indoor scenes. In our project, there are a few approaches carried out to generate output of a scene or a single object. Comparison of the outputs will also be conducted.      

There are challenges and uncontrollable factors while collecting the dataset. Point could collected from LiDAR contains large amount of noise and this could have an impact on the results of surface reconstruction methods. Additionally, with the perfection of known methods, people have more requirements on surface reconstructions as well as map creations. Limitations of the works applied on single objects cannot be utilized for large-scale scenes. Therefore, we desire to discover whether the algorithms applicable for single objects are feasible in large-scale environment creation. Another problem comes along with the increase of dataset. This brings us attention with the efficiency of the algorithms running under larger dataset. Besides comparing the selected algorithms, we will focus on improving the noise effect and time complexity in our future work.

\section{Related Works}


\subsection{Methods for Dense Point Cloud}
Surface Reconstruction methods for Dense Point Cloud has a long history. The main goal is to use a function to express the surface of Point Cloud. Explicit and implicit functions are normally used to achieve this goal.
\paragraph{Traditional Methods}
Hanspeter et,al\cite{surfel} Proposed the method SURFEL for rendering. Surfel is short for surface elements, it contains color, position and other information of the point cloud. It can express the surface explicitly.

SDF method which is short for Signed Distance Function divides the space into voxels.The zero-set of the SDF represents the isosurface of a mesh implicitly. And the Transacted Signed Distance Function (TSDF)\cite{tsdf} is proposed to avoid surface interference. 

In the year 2007, Poisson surface Reconstruction method was proposed\cite{poisson}. The algorithm starts with a set of oriented points which is point clouds with normals. Poisson Surface Reconstruction finds the implicit function by transferring the task into a spatial Poisson problem and it works well for different kinds of point cloud.

Some improvements of Poisson has been made and the algorithm is called Screened Poisson Surface reconstruction\cite{Screened} modifies the octree and multigrid implementation to reduce time complexity and provide more details of the reconstructed surface. 

RBF-base approaches is another way to approximate the mesh based on implicit function. RBF is short for Radial Basis Function and it transforms the reconstruction problem into a multi-variational optimization problem.  However, such a problem is a linear system and it requires lots of time to solve the equations. Moreover, RBF-based methods usually need offset-points that are not easy to find. 

HRBFQI\cite{HRBFQI} inherited properties from RBF-based approaches and utilizes Hermite radial basis functions to solve implicit function. Researchers propose a closed-form formulation to construct HRBF - based implicit function by a quasi-solution. Using quasi-interpolation, HRBF can reduce the time complexity and make reconstruction process even more efficient.

\paragraph{Data-Driven Methods}
Nowadays, Deep Learning has been widely used in various industries and fields. Neural Networks are also adapted to the domain of Surface Reconstruction.  The data driven method are proved to better handle noise and outlier better than the traditional methods.

One of the data drive methods is point2surf\cite{point2surf}. It is a patch-based learning framework that produces accurate surfaces directly from raw data input point cloud. It learns Signed Distance Functions with a designed Neural Network.



Another data driven method Point2Mesh\cite{point2mesh}deforms initial mesh to shrink-wrap a single input point cloud. The local kernels are calculated to fit the overall objects and the method explicitly considers the entire reconstructed shape.

\subsection{Methods for Sparse LiDAR Sequences}
The main difference between the two kinds of method is whether the algorithm is conducting registration in the following methods. When the input is alternated by Sparse LiDAR Sequences, a registration process needs to be done to fuse the spare point cloud of each frame together.

KinectFusion\cite{KinectFusin} is the first method achieve the low - cost hand holding markless surface reconstruction. It is based on the TSDF method, and use rendered map to conduct frame to model ICP.

And the Elastic Fusion\cite{ElasticFusion} proposed a pose-graph free SURFEL based method, it gets rid of the pose graph optimization which is commonly used in Simultaneous localization and mapping (SLAM). It uses a deformable surfel map to conduct registration and reconstruction.

Inspired by the \textit{Elastic Method}, many other similar methods are appeared. The most representative one is \cite{ElasticLidarFusion}. It uses LiDAR as its input sensor. It is quite similar to Elastic Fusion, but optimizes the method for LiDAR and uses IMU constraints.  Kaixuan Wang et al.proposed a method named Real-Time Scalable Dense Surfel Mapping\cite{DeepSurfelMapping}. The method is based on deformable surfel map. It uses superpixel-based surfels to accelerate the running speed but uses existing SLAM part such as vins-mono or orb-slam2 as the odometry input.

In 2018, a new surfel base method short for SuMa\cite{SuMa} was proposed. Unlike the surfel based method mentioned above, it uses a rendered map and its normals to carry out a registration process. While handling map representation, the process still uses surfel. The method is designed for large outdoor environment surface reconstruction.

After SuMa was proposed, Puma\cite{Puma} known as oisson Surface Reconstruction for LiDAR Odometry and Mapping was proposed to handle the outdoor scene. It also uses the implicit method and adds a post-processing step to handle large scenes after Poisson reconstruction was first invented.

However, those method all have their limitations. Elastic LiDAR Fusion treats the system as a continuous system just using the IMU data to help interpolation. Suma and Dense Surfel Mapping do not work well when the input point cloud is sparse.  Puma does not use IMU information to aid Reconstruction.

\section{Replication and Comparison of methods}
Among the algorithms described above, we ran through Poisson, HRBFQI, Point2mesh and PUMA to test their performances and made comparison and contrast among them.

\subsection{Dense Point Cloud}
\subsubsection{Poisson}
Poisson, HRBFQI and Point2mesh are used to reconstruct mesh from a single point cloud data source. Hence, we put them together to compare their performances. Our point cloud data is collected and constructed to show a building and the setting around it. It contains 2,885,708 points with no normal information.

\Figure[htbp!](topskip=0pt, botskip=0pt, midskip=0pt)[scale=0.25]{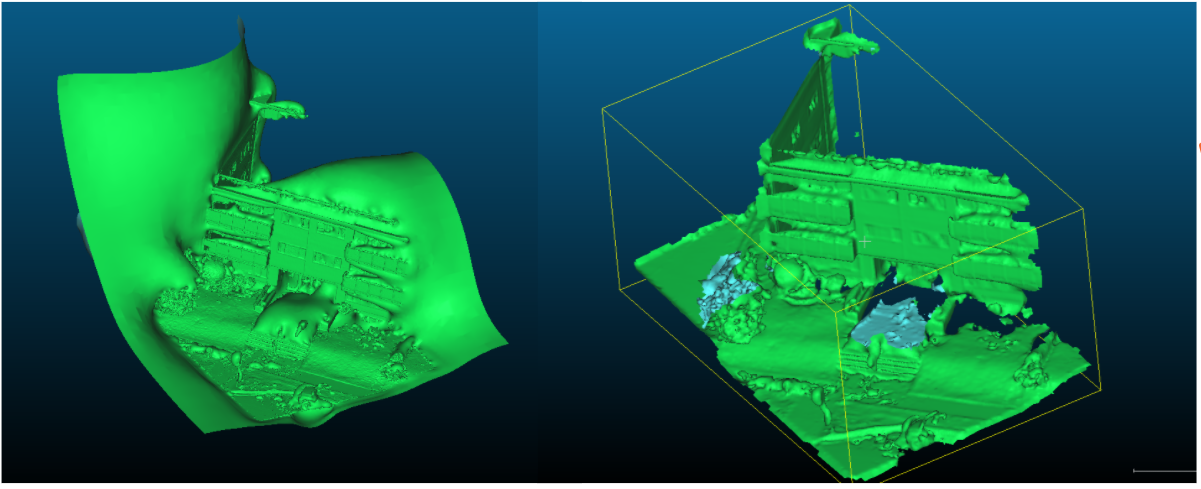}
{pypoisson reconstruction(left) and open3D(right) reconstruction with octree depth 8.\label{pvo}\centering}

First algorithm used was Poisson. We made use of two open sources implementation to see the effect of reconstruction including pypoisson and open3D.  Poisson surface reconstruction needs normal information of point cloud, there we use open3D library to estimate the normal vectors. Using the  depth of an octree 8, figure \ref{pvo} demonstrate the reconstruction effect of pypoisson and open3D. As the figure showed, pypoisson construct unexpected mesh to link all parts of surface together and then create an integrated watertight mesh.  On the other hand, open3D one will keep the information as it is, which is better. Therefore we will use open3D's implementation of Poisson algorithm in the next discussion.

\Figure[htbp!](topskip=0pt, botskip=0pt, midskip=0pt)[scale=0.4]{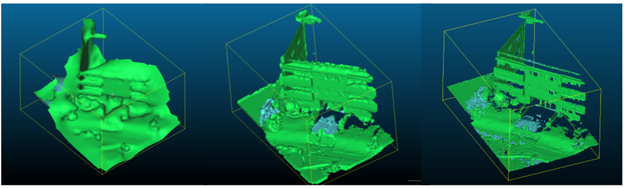}
{open3D's poisson implementation with octree depth 6(left), 8(middle), 10(right)\label{effect_poisson}\centering}

The octree depth has a great impact on the effect of reconstruction. The deeper the octree has, the more detail information will be demonstrated as algorithm works. Figure \ref{effect_poisson} shows the reconstruction effect of open3D's implementation of Poisson algorithm with different octree depth.

However, time complexity should be considered once depth of an octree go deeper. Table \ref{detail_poisson} shows the detailed information while running open3D's implementation of Poisson algorithm.


\begin{table}[htbp]
\caption{Detail of open3D Poisson\centering}
\begin{center}
\scalebox{1.1}
{
\begin{tabular}{ |c|c|c|c|c|} 
 \hline
 Depth of Tree & 6& 8 & 10 & 12 \\ 
 \hline
 Running Time&0.88s&3.91s&35.59s&199.22s \\
 \hline
 Result Triangles &16120&215130&3148301&15593463\\   
 \hline
 Result Points&8178&108856&1596434&7981973\\
 \hline
\end{tabular}
}
\label{detail_poisson}
\end{center}
\end{table}

A surface with more triangular mesh and points would have better visual effect and details, but it also needs more time to process. If we define increase detail factor as
$$D_{increase}=\frac{D_a}{D_b}$$

where $D_a$ means number of resulting triangles mesh with its depth of octree, and $D_b$ means number of resulting triangles with shallow octree. $D_{increase}$ explains how much detail it will performance by increasing the depth of the octree. Therefore we define increase time factor as
$$T_{increase}=\frac{T_a}{T_b}$$

here $T_a$ refers to time for processing with deep octree, and $T_b$ shows time for processing with shallow octree. $T_{increase}$ explains how much time cost it gets by increasing depth of tree.

From functions $D_{increase}$ and $T_{increase}$, we have noticed that more details of surface increasing, it means we need a deeper depth of an octree and a cost more time to run the process.

$$E = \frac{D_{increase}}{T_{increase}}$$ In this function,  when $E$ is greater than 1, we can tell we are getting better results with increasing the depth of an octree. Once $E$ is smaller than 1, it's not beneficial to increase the octree's depth and that means we need more time to process with more details. With the detailed information showed in figure 3, it shows that when depth of an octree is lower than 10, we have $E$ greater than 1. That means we are beneficial with increasing the octree's depth. Therefore, the preferable octree's depth should be 10.

\subsubsection{HRBFQI}
On top of Poisson, we also implement surface reconstruction using HRBFQI. It depends on a quasi-solution equation and needs two parameters.  One parameter is scalar to the initial support size computed with the given data short for SSS.  Another parameter is the size of the grid edge for isosurface extraction short for IES. 

Table \ref{detail_SSS} and Figure \ref{effect_SSS} show the details of HRBFQI with different SSS and the effect of result.


\Figure[htbp!](topskip=0pt, botskip=0pt, midskip=0pt)[scale=0.3]{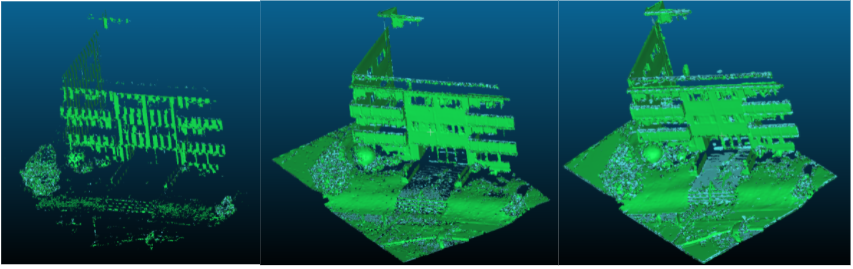}
{Effect of Result with Different SSS 1.2(Left) 2.4(Middle) 4.8(right)\label{effect_SSS}\centering}


\begin{table}
\caption{Detail of HRBFQI with Different SSS \centering}
\begin{center}
\scalebox{1.2}
{
\begin{tabular}{ |c|c|c|c|} 
 \hline
 Value of SSS& 1.2& 2.4 & 4.8 \\ 
 \hline
 Running Time&10s&17s&43s \\
 \hline
 Result Triangles &107914&305072&420362\\   
 \hline
 Result Points&90671&162276&223618\\
 \hline
\end{tabular}
}
\label{detail_SSS}
\end{center}
\end{table}

\Figure[htbp!](topskip=0pt, botskip=0pt, midskip=0pt)[scale=0.3]{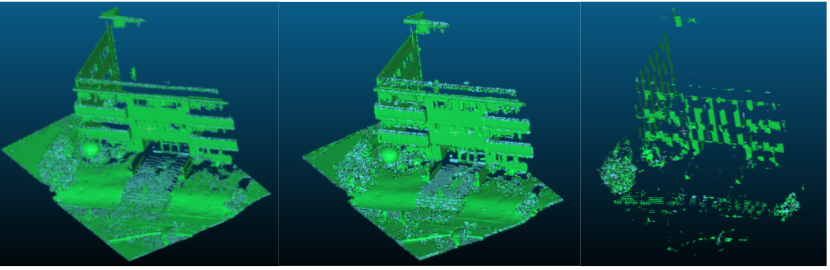}
{Effect of result with different IES 0.05(left) 0.1(middle) 0.2(right)\label{effect_IES}\centering}
\begin{table}
\caption{Detail of HRBFQI with Different IES \centering}
\begin{center}
\scalebox{1.2}
{
\begin{tabular}{ |c|c|c|c|} 
 \hline
 Value of IES& 0.05& 0.1& 0.2 \\ 
 \hline
 Running Time&61.47s&17s&7.79s \\
 \hline
 Result Triangles &1599148&305072&30112\\   
 \hline
 Result Points&844207&162276&24158\\
 \hline
\end{tabular}
}
\label{detail_IES}
\end{center}
\end{table}

As the figures shown, HRBFQI will produce a more detailed mesh when SSS increases. Same as Poisson, we can calculate $E$ to see if we are getting benefit when increasing SSS. The preferable SSS is 2.4

Table \ref{detail_IES} and figure \ref{effect_IES} show the detail of HRBFQI with different IES and the effect of result.

On top of that, we also noticed HRBFQI will produce a more detailed mesh when IES decrease. Similarly, we can calculate $E$ to see if we are getting benefit when decreasing IES. The preferable IES is 0.05


\subsubsection{Points2Surf}
Points2Surf is a data driven method that is used to reconstruct the surface. It needs a lot of dense point cloud data to train the model, and requires longer time than Poisson and HRBFQI. We made a comparison of time between different methods therefore We can conclude Poisson is the fastest algorithm among the three of them to reconstruct a surface of dense point cloud data.


\begin{table}
\caption{comparison of different methods \centering}
\begin{center}
\scalebox{1.2}
{
\begin{tabular}{ |c|c|c|} 
 \hline
 Result Triangles Range & 10K-30K& 200K-300K \\ 
 \hline
 Poisson &1-2s&35s \\
 \hline
 HRBF  &>10s&-\\   
 \hline
 Point2Surf &60s&-\\
 \hline
\end{tabular}
}
\end{center}
\end{table}

\subsection{For Sparse LiDAR Sequences}
We replicate PUMA for sparse LiDAR sequences. It is based on Poisson surface reconstruction, but it also adds features like registering point cloud data within a time period and contains a post processing on reconstructed mesh. We use two dataset to show the result of PUMA.  The first one is the KITTI odometry dataset, which is a benchmark for SLAM evaluation. The data is collected with velodyne LiDAR and there are 22 sequences of LiDAR from the datset. The result of PUMA running on sequence 20 is showed in figure \ref{puma_kitti}.

\Figure[t!](topskip=0pt, botskip=0pt, midskip=0pt)[scale=0.2]{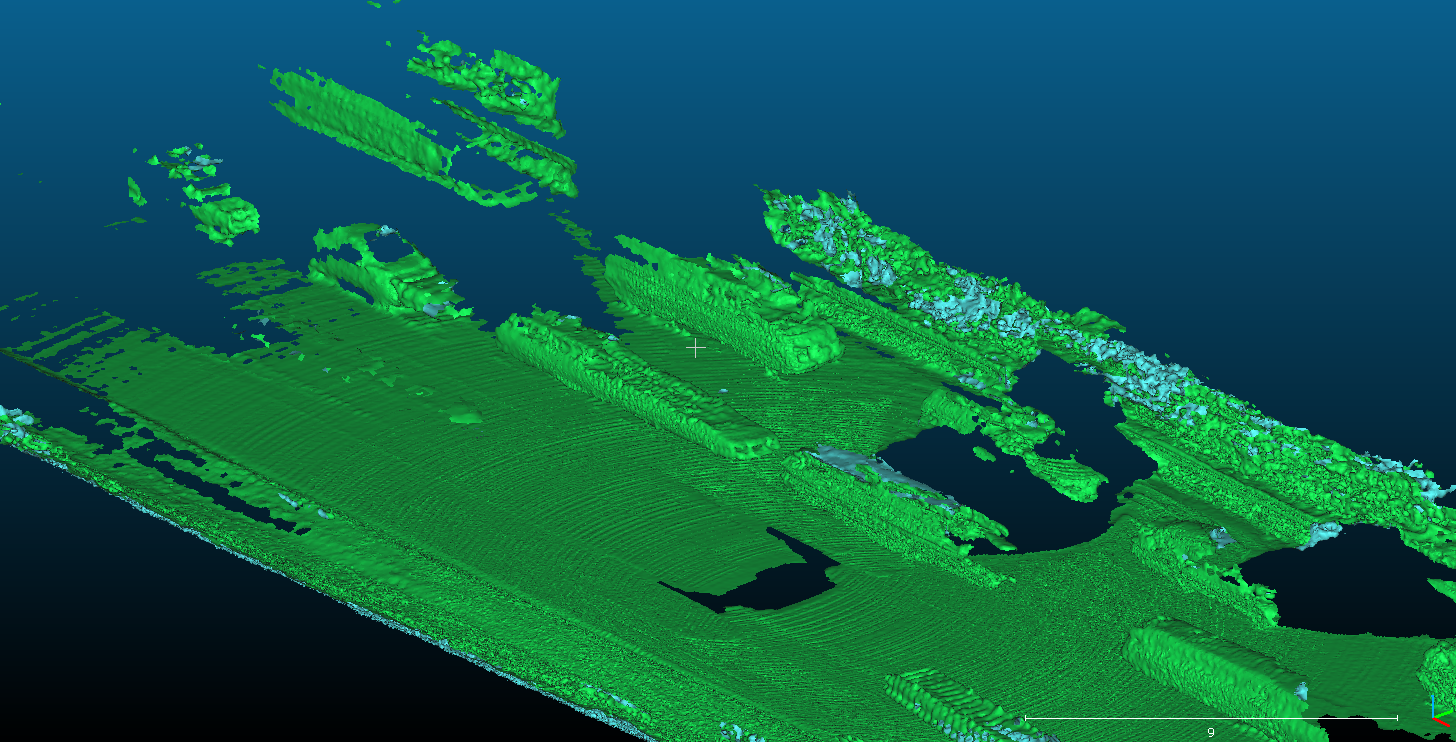}
{PUMA result on KITTI sequence 20\label{puma_kitti}\centering}

Another dataset used in our project is Mai - city which contains 3 different data sequences. The data was collected by placing virtual sensors on the 3D cad model and, through ray-casting techniques, a LiDAR scan is obtained by sampling the original mesh. Figure \ref{puma_mai} shows the result of PUMA running on sequence 01.

\Figure[t!](topskip=0pt, botskip=0pt, midskip=0pt)[scale=0.2]{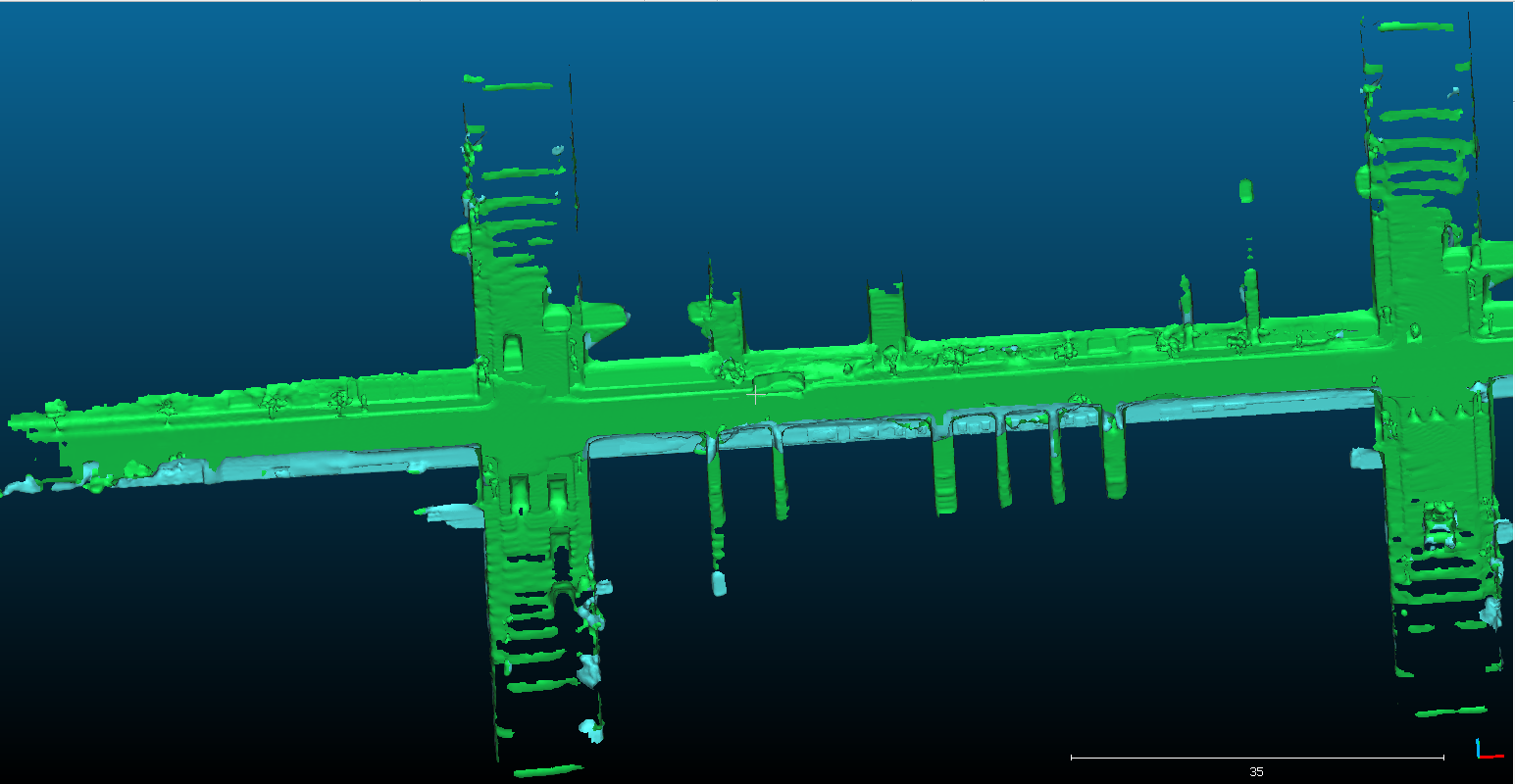}
{PUMA result on Mai city Sequence 01\label{puma_mai}\centering}

From our output, we see the deficiency of puma for moving objects. If we directly use those data to do reconstruction, the resulting mesh will also be distorted. Therefore, finding the distorted part of point cloud and removing it would potentially be a future work of us.

\section{Proposed Method}

\subsection{Deskewed point cloud and odometry Generation}

We use SLAM as a source of deskewed point cloud and odometry. Then LIO - SAM is found suitable for our project.  Therefore, we modify it to adapt to the whole system. LIO-SAM is a kind of tightly-coupled-system.

\subsubsection{Point Cloud Motion Compensate}
It uses the IMU data to compensate the distortion of the LiDAR point cloud due to the motion in a round of scan.

\subsubsection{LiDAR Point CLoud Registration}
Similar to LOAM\cite{LOAM}, The each point cloud position in the world frame is optimized by minimizing the residual measured by the edge residual and the plane residual. Using Gauss-Newton Method, we need to solve the equation
$$J^TJ\cdot \Delta T=-J^Tf$$ where $f$ is the residual.

$f$ is the combination of corresponding edge or plane residual. Using $d_e$ and $d_g$ to represent them. 

Edge residual represents the distance between the edge point detect in the $k+1$ frame transformed to the world frame, and the corresponding scan line in the $k$ frame. 
$$d_e=\frac{\vert\tilde{x}^W_{k+1,i} -\bar{x}^W_{k+1,j}\vert\times\vert\tilde{x}^W_{k+1,i} -\bar{x}^W_{k+1,l}\vert}{\vert \bar{x}_{k,j}^W-\bar{x}_{k,l}^W\vert}$$

The plane residual is the error between the transformed point the its corresponding plane.

$$d_H=\frac{pa\cdot x+pb \cdot y +pc\cdot z+pd}{\sqrt{pa^2+pb^2+pc^2}}$$

The goal is $$min\left\{ \sum_{i\in n_e}d_{e_i}+\sum_{k\in n_d}d_{p_k} \right\}$$
Where $n_e$ and $n_d$ represents the set for matched edge feature points and plane feature points in the two continuous frame.

\subsubsection{Pose Graph Optimization}
The poses of the key frames will be optimized while marginalization is done to maintain the fixed size of the $H$ Matrix. It is impossible to store all the constraints for all the poses. We need to obsolete those old frames and retain the some of the constrains by updating the $H$ matrix using Schur Complement.

To solve an existing optimization problem, we can use Schur Complement to solve $X_r$ without using $X_m$
$$\left[
 \begin{matrix}
   H_{mm}& H_{mr} \\
   H_{rm}&H_{rr} \\
  \end{matrix}\right]\left[
 \begin{matrix}
   X_{m} \\
   X_{r} \\
  \end{matrix}\right]=\left[
 \begin{matrix}
   b_{m}\\
   b_{r} \\
  \end{matrix}\right]$$
  
Then the optimization problem will be decomposed into 

  \begin{equation*}
  \left[
 \begin{matrix}
   H_{mm}& H_{mr}\\
   0&H_{rr}-H_{rm}H_{mm}^{-1}H_{mr} \\
  \end{matrix}\right]
  \left[
 \begin{matrix}
   X_m\\
   X_r\\
  \end{matrix}\right]=
  \left[
 \begin{matrix}
   b_m\\
   b_r-H_{rm}H_{mm}^{-1}b_m\\
  \end{matrix}\right] \end{equation*}
And we can solve the $x_r$ without $x_m$ by using the equation
$$(H_{rr}-H_{rm}H_{mm}^{-1}H_{mr})X_r=b_r-H_{rm}H_{mm}^{-1}b_m$$

This step above is named Marginalization.

\subsubsection{IMU - preintegration}
IMU - preintegration is conducted at the same time when Pose Graph Optimization was invented. IMU can measure the movement of an object in a high frequency. However, IMU can only measure the deviation of the movement such as the velocity and angular velocity.

To get the accurate pose information consists of position and orientation, we need to solve the differential equation and the integration must be done. We need the equation below to obtain the position, velocity and rotation. 

$$P_{wb_j}=P_{wb_i}+v_i^w\Delta t+\iint_{t\in[i,j]}(q_{wb_t}a^{b_t}-g^w)\delta t^2$$
$$v_j^w=v_i^w+\int_{t\in[i,j]}(q_{wb_t}a^{b_t}-g^w)\delta t$$
$$q_{wb_j}=\int_{t\in[i,j]}q_{wb_t}\otimes\left[\begin{matrix}
  0\\
  \frac{1}{2}\omega^{b_t}
\end{matrix}\right]\delta t$$


The pose of the object may be updated in optimization process.  Then the pre - integration technique has been invented\cite{manifold}.

By alternating $q_{wb_t}=q_{wb_i}\otimes q_{b_ib_t}$, we can get the pre-integration equation.
$$p_{w_j}=p_{w_i}+v_i^w\Delta t-\frac{1}{2}g^w\Delta t^2+q_{wb_i}\underbrace{\iint_{t\in[i,j]}(q_{b_ib_t}a^{b_t})\delta t^2}_{\alpha_{b_ib_j}}$$
$$v_j^w=v_i^w-g^w\Delta t+q_{wb_i}\underbrace{\int_{t\in[i,j]}(q_{b_ib_t}a^{b_t})\delta t}_{\beta_{b_ib_j}}$$
$$q_{wb_j}=q_{wb_i}\underbrace{\int_{t\in[i,j]}q_{b_ib_t}\otimes\left[\begin{matrix}
  0\\
  \frac{1}{2}\omega^{b_t}
\end{matrix}\right]\delta t}_{q_{b_ib_j}}$$

The $\alpha_{b_ib_j}$,$\beta_{b_ib_j}$,$q_{b_ib_j}$ are the pre-integration, which can get rid of the pose $i$ in the world  frame.

\subsubsection{Pose Graph Optimization}
The optimized pose of the keyframes will be send to correct the IMU bias and update the IMU trajectory. 

In our SLAM part, we utilized the LiDAR registration as the prior knowledge for to optimize the pose for example the point cloud in spatial coordinates.

We also use $se(3)$ to represent homogeneous transform including the rotation and translate.
Error between two poses $e_{ij}$is define by $$ln(T_{ij}^{-1}T_i^{-1}T_j)^\vee=ln(exp((-\xi_{ij})^\wedge)exp((-\xi_i)^\wedge)exp(\xi_j)^\wedge)^\vee$$

To use the non-linear optimization method mentioned above, the Jacobian Matrix is required. We suppose the transform has been obtained, such as using LiDAR point cloud registration. We need to optimize the poses of each frame to obtain a more accurate result. 
We define the $$\hat{e_{ij}}=ln(T_{ij}^{-1}T_i^{-1}exp((-\delta\xi_i)^\wedge)exp(\delta\xi_j^\wedge)T_j)^\vee$$

Using the ad-joint property and BCH function, we can obtain that $$\hat{e_{ij}}\approx e_{ij}-\underbrace{\mathcal{J}_r^{-1}(e_{ij})Ad(T_j^{-1})}_{A_{ij}}\delta\xi_i+\underbrace{\mathcal{J}_r^{-1}(e_{ij})Ad(T_j^{-1})}_{B_{ij}}\delta\xi_j$$
Where $$\mathcal{J}_r^{-1}(e_{ij})\approx I+\frac{1}{2}\left[ \begin{matrix}\phi_e^\wedge&\rho_e^\wedge\\0& \phi_e^\wedge\end{matrix}\right]$$
Then we can calculate the Jacobian of the matrix, which can be used to optimize the poses in the SLAM system. $$J_{ij}=(0\dotsb \underbrace{A_{ij}}_{pose_i}0\dotsb0\underbrace{B_{ij}}_{pose_j}0)$$

  For a matrix $M$,it can be transformed into 


\subsection{Voxgraph}
The pipeline for TSDF based method is as follows. When a frame of data in the LiDAR has been accepted in the system.The Voxgrpah will update the TSDF corresponding to the voxel with assigned weights. The first method conduct real-time low-cost surface reconstruction is KinectFusion\cite{KinectFusin}. 
There is a series of works recently on TSDF published based method have been published by the ASL group in ETH Zürich, such as Voxblox\cite{voxblox},Cblox\cite{cblox} and Voxgraph\cite{voxgraph}. Those works use the TSDF method to generate the mesh of an outdoor environment.

In Voxblox, researchers propose a $w_{quad}$ to alternate the fixed weight in the traditional TSDF problem. Cblox optimizes the processng part for LiDAR while the main idea of voxgraphh is to generate several sub-maps with continuous sequences of input LiDAR data. Then it will concatenate those sub-maps together by minimizing the transform error and to generate the whole TSDF map.

Then we could use Voxgraph as the recipient of the deskewed point cloud and odometry.

\subsection{Robot Operating System}
Our project was run on the Robot Operating System short for ROS in Ubuntu System. It provides several basic packages for the robot developers. The most important part in ROS is its distributed messaging mechanism. 

In ROS, there are different hierarchies for programs. The basic one is the \textit{node}. We can divide a task into several parts and implement them in each node and these nodes constitute the entire package. ROS has provided several basic message types for robot navigation which can be used to pass the point cloud and odometry.

Developers could use the \textit{Advertise} and \textit{Subscribe} method and pass the wrapped message between nodes and packages. In the SLAM system, the transformation relationship is the most important part. We can use the \textit{TF} package to broadcast the transformation between coordinates and track the pose changes of an object especially the point cloud in the SLAM system.

\Figure[t!](topskip=0pt, botskip=0pt, midskip=0pt)[scale=0.25]{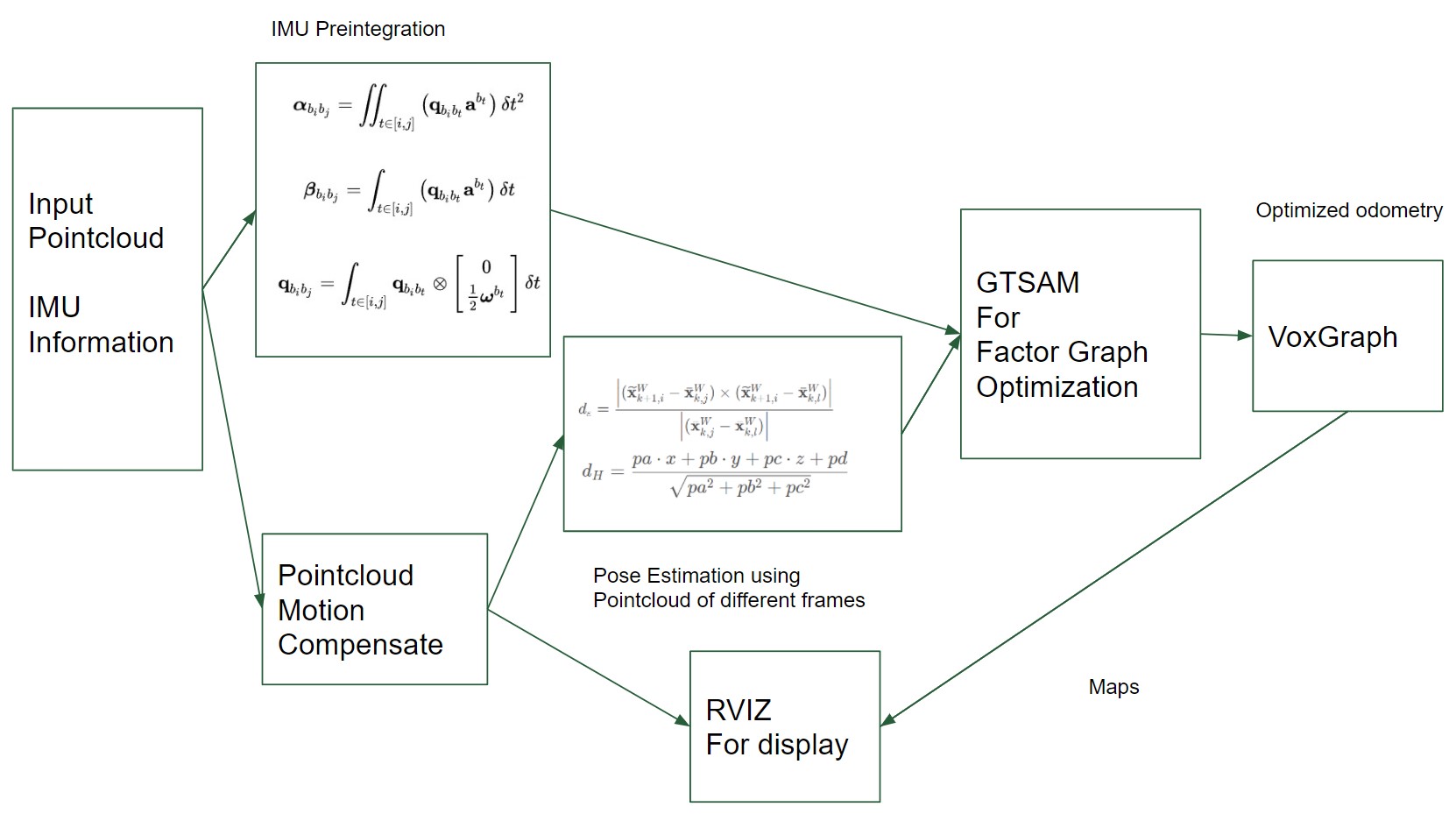}
{Diagram of the SLAM and Voxgraph.\label{diagram} \centering}

\section{Contribution}
With the utilization of ROS, we modify the code in  LIO-SAM and Voxgraph and combine them togethr, then the Voxgraph can handle the LiDAR data without ground truth. We modify the LIO-SAM as a part of SLAM to acquire the the position and the orientation information of the point cloud as the input of the Voxgraph. Motion compensation is also conducted to correct the deskew of point cloud due to the motion. 

As we involve the IMU information as the constraint in the SLAM and it is tightly coupled with the LiDAR odometry, the poses of point cloud in the world frame will be optimized to a more accurate result. A more precise odometry means we can have a reconstructed surface with a higher accuracy. Our system is more flexible and more suitable for different kinds of environment.

We can not finish the entire reconstruction of the dataset, but we can also reconstruct a unknown environment using the LiDAR on a robot, and provide the 3D map of the environment to achieve the obstacle avoidance.

\Figure[t!](topskip=0pt, botskip=0pt, midskip=0pt)[scale=0.2]{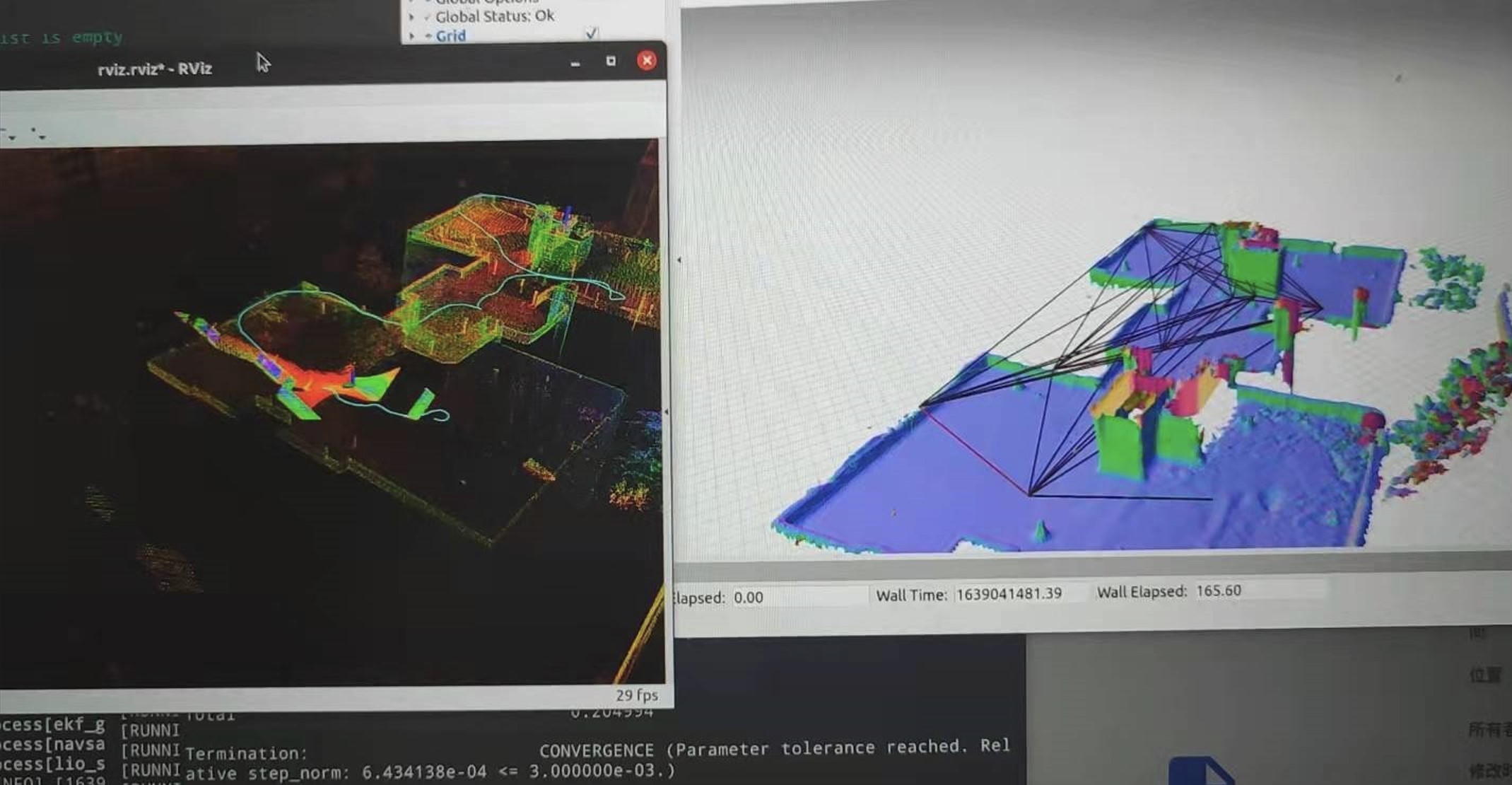}
{Screenshot of the system.\label{sreenshot} \centering}

\section{Future Work}
In the SLAM part, the IMU inforamtion is guided by the optimized map, we will try to optimize the map with IMU and LiDAR constraints together. 

From the existing Dense Monocular Surface Reconstruction works, we can see that neural network can be used as a feature extractor of an input image, however, rare works has been done on LiDAR point cloud feature extraction.We could train a neural network based base feature extractor to alternate the geometry based feature extractor. In this way, we may find out how to detect moving objects in time sequences and modify them to avoid mesh distortion.

\printbibliography


\EOD

\end{document}